\documentclass{elsart}
\usepackage{amssymb}
\usepackage[dvips,unicode]{hyperref}
\usepackage[dvips]{graphicx}

\begin{document}

\begin{frontmatter}

\title{The measurements, parameters and construction of
Web proxy cache}

\author{Dmitry Dolgikh}

\address{Samara State Aerospace University, Moscovskoe sh. 34a,
Samara, 443086, Russia}

\author{Andrei Sukhov\corauthref{avt} }
\address{Laboratory of Network Technologies, Samara Academy
of Transport Engineering, 1 Bezymyanny per.,
18, Samara, 443066, Russia}

\thanks[avt]{Corresponding author\\{\em E-mail addresses:
sukhov@ssau.ru}~(Andrei M. Sukhov), {\em ddolgikh@ssau.ru}~(Dmitry
G. Dolgikh)}

\begin{abstract}
The aim of this paper is an experimental study of cache systems in
order to optimize proxy cache systems and to modernize
construction principles. Our investigations lead to the criteria
for the optimal use of storage capacity and allow the description
of the basic effects of the ratio between construction parts,
steady-state performance, optimal size, etc. We want to outline
that the results obtained and the plan of the experiment follow
from the theoretical model. Special consideration is given to the
modification of the key formulas supposed by Wolman at
al.~\cite{wolm}.
\end{abstract}

\begin{keyword}
Cache system \sep Zipf-like distribution \sep document lifetime
\sep elements of cache construction \sep renewal of Web documents
\end{keyword}
\end{frontmatter}

\section{Introduction}
\label{Intr}

This paper considers approaches to the problem of optimization of
proxy cache construction based on theoretical and experimental
study. Until recently, caching was an optional service for users
who voluntary configured their browsers to redirect request
through a proxy. The Internet Service Providers interpose the
caching systems in the strategic places at the organization
boundaries.

In experimental research of the cache systems and construction of
the mathematical models should lead to the growth of the caching
effectiveness with minimal financial expenditures. A better
algorithm that increases hit ratios by only several percent would
be equivalent to a multiple growth in cache size. In order to find
the optimal cache size Kelly and Reeves~\cite{kelr} are guided on
economical methods like monetary cost of memory and bandwidth.

Our analysis is based on the model presented by Breslau et
al.~\cite{bres} and extended by Wolman et al.~\cite{wolm} to
incorporate the steady-state behavior and documents' rate of
change. One difficulty in the study of Web caching is that there
are many cache replacement policies and many factors affecting
their performance. Wolman et al. parameterized the model using
population size, population request rate, document rate of change,
size of object universe, and popularity distribution of objects.
Several researchers~\cite{jinb} include additional factors like
object size, miss penalty, temporal locality, and long-term access
frequency. Our research allows the calculation of the lower limit
of the cache size corresponding to aggregated bandwidth of
external links when performance mount to the effective level equal
to 35\%.

Such an approach can be easily generalized to describe any
applications based on Zipf-like distribution such as Content
Distribution Networks (CDN)~\cite{gadcr}, peer-to-peer
systems~\cite{rfi,srip}, Internet search engines, etc.

The paper investigates how the different parameters of proxy cache
influence on its performance. The methods of the measurements
including the treatment of experimental data and analytical
formulas for calculations are discussed. The correlations between
the significant parameters are investigated and the corresponding
Figures and Tables are constructed. On the basis of the new
experimental data the analytical model is modernized and the
addition to the cache construction and to replacement algorithm is
proposed.

Special consideration is given to the modification of the key
formulas introduced by Wolman at al.~\cite{wolm}. In order to
describe the renewal effect of Web documents in the global network
the alternative model is developed. The document rate of change
$\mu(i)$ is supposed to depend on popularity index $i$ so the
Zipf-like distribution with the new exponent $\alpha_R$ describes
the mentioned effects of the Wolman at al. model. As a result of
the special experiment the values of exponents $\alpha$,
$\alpha_R$ and of the document rates of change $\mu_p$, $\mu_u$
are calculated.

\section{Plan of the measurements }
\label{exper}

\begin{figure}
\begin{center}
\includegraphics[width=0.45\textwidth]{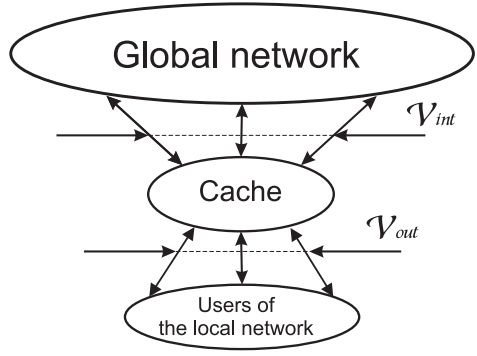}
\end{center}
\caption{Scheme of the proxy cache} \label{scheme-e}
\end{figure}

The scheme of the Web caching could be presented in the following
way: there are users which send requests to the global network and
receive information through a cache system as it is shown on the
Fig.~\ref{scheme-e}. Some documents are requested repeatedly and
therefore they should be held in the cache system.

The relative frequency~\cite{bres} of requests to Web pages
follows Zipf-like distribution~\cite{zipf}. This distribution
states that the relative probability of a request for the $i$'th
most popular page is
\begin{equation}
\label{zipf-gen}
  \theta_i=\frac{A}{i^\alpha},
\end{equation}
where $A=\theta_1$ is the probability of the most popular item and
$\alpha$ is a positive exponential value less then unity.

We proponed the following test: the size of proxy cache $S_{eff}$
was varying. These values correspond to incoming traffic for one,
two, three, and six days. All types of documents both cacheable
and uncacheble are taken into account for calculating the ratio
$S_{eff}/\nu_{int}$.

The network of Samara State Aerospace University has been chosen
as an experimental field. The proxy cache of SSAU is a
two-processor Linux server with SQUID proxy installed. All
hierarchical links were disconnected before the experiment began.
The statistics of requests for the long time $T_{st}\geq  month$
were collected for each point.

In order to modify the model proposed by Wolman at al. we
collected the requests during the time $T_{st} \lesssim  t_u$. The
$t_u$ is the mean lifetime of those documents, who's popularity is
$\vartheta_i=1$ ($\vartheta_i=\theta_i k$).

\section{The original results and their processing}
\label{inres}

The primary results of experiments are summarized in the
Table~\ref{PR}, where the following notations and abridgements are
used:
\begin{itemize}
  \item
The variables $\nu_{int}$ and $\nu_{out}$ are the incoming and the
outgoing request's streams of cache proxy as it is shown on the
Fig.~\ref{scheme-e}. They are measured as the number of users
requests per day (Rpd - request per day). It should be noted that
the variable $\nu_{out}=\lambda N$ describes the request stream
from a collective user that is a significant parameter of the
Wolman at al. model.
\item
$H$ is the performance of the cache system or, in other words, hit
ratio.
\item
The same variables with upper index $B$ describe the system in the
units of transmitting traffic ($Kbps$ - Kbit per second).
\item
$E(S)$ is the mean size of documents received from the global
network directly.
\item
$E(C)$ is the mean size of documents from the cache.
\item
Finally, the time $T_{st}$ corresponds to the quantity of days
when statistics are collected.
\end{itemize}

\begin{table*}
\caption{Primary results}\label{PR} {\scriptsize
\begin{tabular}{|c|c|c|c|c|c|c|c|c|c|}
\hline $\frac{S_{eff}}{\nu_{int}}$ & $\nu_{out}$ & $\nu_{int}$ &
$H$ & $\nu^B_{out}$ & $\nu^B_{int}$ & $H^B$ & $E(C)$ & $E(S)$ &
$T_{st}$
\\days & $\times10^3$ & $\times10^3$ & \% & $Kbps$ & $Kbps$ & \% &
$Kbyte$ & $Kbyte$ & days
\\ & $Rpd$ & $Rpd$ &&&&&&& \\
\hline 1.0 & 56.5 & 42.5 & 24.49 &  42.6 & 38.6 & 9.13 & 8.13 &
10.5 & 31\\ \hline 2.15 & 53.4 & 39.5 & 28.08 & 44.1 & 40.3 &
10.33 & 8.9 & 12.5 & 28\\ \hline 3.15 &69.8 & 47.3 & 32.19 & 46.8
& 41.5 & 11.17 & 7.25 & 13.7 & 31
\\ \hline 5.96 & 69.8 & 42.3 & 36.75 & 56.3 & 50.2 & 10.78 & 8.71 & 13.8 & 61 \\ \hline
\end{tabular}}
\end{table*}

The statistics collected were processed by scripts specially
written for the task. Originally, so-called cacheable documents
that can be stored in a proxy are selected from the general list.
Later the corresponding Zipf-like distribution was constructed
where the documents were placed in the order of reducing of
popularity index $\vartheta_i$. The fragment of this list is shown
bellow:

{\scriptsize
\begin{tabular}{lll}
112 & http://www.ixbt.com/images/empty.gif & (line 457), i.e.
$\vartheta_{457}=112$\\ 111 &
http://cacheserver.myecom.net/main/images/adlogo.jpg  & (line
458)\\ .&.&.\\ 2 & http://zzz.net.ru/images/spacer.gif & (line
78166=$M$)\\.&.&.\\ 1 & http://www.muz-tv.ru/chat/chat-top.html
&(line 200045)
\end{tabular}}

The number of unique cacheable documents or the quantity of the
lines in the list mentioned above is $p$. The line number of the
last document, which was requested twice ($\vartheta_M=2$), is
$M$.

In order to find the general number of cacheable documents $k$ the
following sum was calculated
\begin{equation}\label{k-gen}
  k=\sum\limits_{i=1}^p  \vartheta_i
\end{equation}
where $ \vartheta_i$ is the number of cache requests for the
$i$'th most popular document. The portion of cacheable documents
$p_c$ is defined as
\begin{equation}\label{p-c}
  k=p_c\nu_{out}T_{st},
\end{equation}
where $K=\nu_{out}T_{st}$ is the general number of all documents
received from the global network including both cacheable and
uncacheable ones.

The value of $\alpha$'s was calculated using the equation
\begin{equation} \label{alpha-p}
  \alpha=1-2M/\sum\limits_{i=1}^M\vartheta_i=1-2M/(k-p+M)
\end{equation}

Finally, analyzing the {\em log\/} files mean lifetime $t_u$ was
calculated for those documents, with a popularity of
$\vartheta_i=1$. Such statistics also determine the lifetime
$T_{eff}$ of cache objects with the citing index $\vartheta_i=2$,
i.e. those items, which have been stored in proxy cache, requested
one time from a proxy, and deleted subsequently (see
Tab.~\ref{SR}).
\begin{table*}
\caption{Parameters of proxy cache}\label{SR}{\scriptsize
\begin{tabular}{|c|c|c|c|c|c|c|c|c|c|c|}
\hline $\frac{S_{eff}}{\nu_{int}}$ & $S_{eff}$ & $\alpha$ & $t_u$
& $T_{eff}$ & $p_c$ & $M$ & $p$ &$k$
\\days & {\scriptsize $\times10^5$} && {\scriptsize days}&{\scriptsize days}  &
&{\scriptsize $\times10^5$} & {\scriptsize $\times 10^5$}
&{\scriptsize $\times10^5$}
\\  \hline 1.0 & 0.42 & 0.76 & $2.2\pm0.9$ & $3.8\pm1.9$ & 0.59 &0.99& 3.10& 10.4\\ \hline
2.15 &  0.86 & 0.77 & $6.8\pm1.9$ & $9.1\pm4.6$ & 0.58 & 0.78 &
2.48 & 8.7\\ \hline 3.15 & 1.49 & 0.74 & $8.8 \pm 2.9$ &
$8.5\pm3.6$ &0.56& 1.22 & 3.68 & 12.0
\\ \hline
5.96 & 2.53 & 0.81 & $20.4\pm 2.4$ & $18.9\pm 7.0$ & 0.59  & 2.01
& 6.07 & 25.0
\\ \hline
\end{tabular}}
\end{table*}

\section{Basic correlation}
\label{correl}

\begin{figure}
\begin{center}
\includegraphics[width=0.45\textwidth]{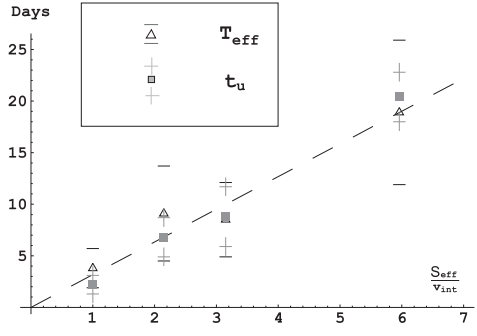}
\end{center}
\caption{Dependence between lifetimes and cache size} \label{Tt}
\end{figure}

The first family of the curves, which should be analyzed, is the
dependence of $t_u$ and $T_{eff}$ on cache size
$S_{eff}/\nu_{int}$. As it was discovered in the
Ref.~\cite{dols1}, these curves define the ratios between elements
of cache construction:
\begin{itemize}
\item
A kernel $S_k$ that contains popular documents with
$\vartheta_i\geq 2$.
\item
An accessory part $S_u$ that keeps unpopular documents requested
from the Internet once, i.e. $\vartheta_i=1$.
\item
A managing part $S_m$ that contains statistics of requests and
rules for replacement of cache objects.
\end{itemize}
Earlier~\cite{dols,dols1} we have got the following Egs.
\begin{eqnarray}
S_k &= & \frac{(1-\alpha)H}{2}\nu_{out}T_{eff} \label{S_k}\\
\label{sk/su-n}
 \frac{S_k}{S_u}&=&\frac{T_{eff}}{(2^{1/\alpha}-1)t_u}=\frac
 M{p-M}\frac{T_{eff}}{t_u}
\end{eqnarray}
Analysis of the experimental data shown in the Tab.~\ref{SR} and
on the Fig.~\ref{Tt} leads to the facts that variables $t_u$ and
$T_{eff}$ are directly proportional to the cache size
$S_{eff}/\nu_{int}$ and can be considered as coincided values:
\begin{equation}
\label{t=T} T_{eff}\simeq t_u
\end{equation}
The only deflection indicated for effectiveness of replacement
algorithms is revealed at small cache size when performance is far
from the optimal value.

In other words the kernel and accessory parts are approximately
correlated as 1:2 or less then 40\% of storage capacity has been
used for the basic goal to store the repeatedly requested
documents. It is remarkable that the number of documents $M$,
which must be stored in the cache system for one and two months
traffic, is less then three and six-days incoming stream
correspondingly.

\begin{figure}
\begin{center}
\includegraphics[width=0.45\textwidth]{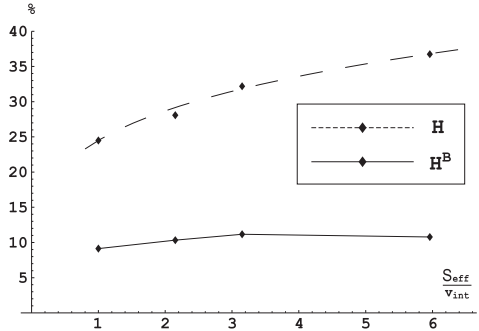}
\end{center}
\caption{Dependence between hit ratios and cache size} \label{HHb}
\end{figure}

The second family of curves intended for study is the dependence
of the system performance $H$ and $H^B$ on its relative size
$S_{eff}/\nu_{int}$, see Fig~\ref{HHb}.

The hit rate $H$ of the web cache is considered to grow in a
log-like fashion as a function of cache size \cite{alm,bres,caoi}.
From the expression for cache performance
\begin{equation}\label{H-r}
 H=p_c\int\limits_1^{S_k}\frac{A}{x^\alpha}dx,
\end{equation}
the following dependence appears
\begin{equation}\label{H-S}
  \frac{H_1}{H_2}=\left( \frac{S_1}{S_2}\right)^{1-\alpha},
\end{equation}
that allows us to talk about power fashion.

The Fig~\ref{HHb} illustrates the fact that the dependence between
$H$ and $S_{eff}/\nu_{int}$ is successfully described by
Eq.~(\ref{H-S}) with $\alpha=0.77$ and the curve $H^B$ is not
predictable. This effect needs additional study especially because
the Tab.~\ref{PR} shows a positive difference between the mean
size of cacheable documents $E(S)$ and the mean size of all items
$E(C)$.

\section{Renewal of Web documents}
\label{renew}

Rapid development of computer technology at the end of the last
millennium lead to the appearance of a virtual world with its own
laws. Unfortunately, during this period little attention was given
to studying of fundamental principles of virtual life.

Nowadays we can afford some respite and some resources should be
transferred to studying of the fundamental laws and to
optimization of the vital systems on the basis of recent
knowledge. Frankly speaking, any research area may be considered
as a scientific field if and only if its basic principles are
enveloped in mathematical form and new rules can be predicted on
the basis of confirmed facts.

Our analysis is based on the model presented by Breslay et
al~\cite{bres} and extended by Wolman et al.~\cite{wolm} to
incorporate document rate of change. The Wolman model yields
formulas to predict steady-state properties of Web caching systems
parameterized by population size $N$, population request rates
$\lambda$, document rate of change $\mu$, size of object universe
$n$ and a popularity distribution for objects.

The key principles of their theoretical study has been developed
in our work~\cite{dols1} to describe the basic effects as a ratio
between construction parts, steady-state performance, optimal
size, etc. in an alternative way. We want to outline that the
results received and the plan of the experiment follow from the
theoretical model~\cite{dols1}.

The key formulas from Wolman et al.
\begin{eqnarray}
C_N
&=&\int\limits_1^n\frac{1}{Cx^\alpha}\left(\frac{1}{1+\frac{\mu
Cx^\alpha}{\lambda N}}\right) dx \label{C_N}\\ C
&=&\int\limits_1^{n}\frac{1}{x^{\alpha}} dx \label{C-old}
\end{eqnarray}
yield $C_N$, the aggregate object hit ratio considering only
cacheable objects. Document rates of change $\mu$ were considered
to take two different values, one for popular documents $\mu_p$
and another for unpopular documents $\mu_u$. An additional
multiplier from Eq.~(\ref{C_N}) throws off one effective query to
any document from cache during time $T_{ch}=1/\mu$ between its
changes. Such an updating request must be redirected to the global
network to get the renewed Web page.

\begin{figure}
\begin{center}
\includegraphics[width=0.45\textwidth]{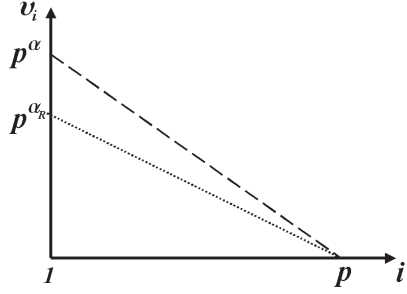}
\end{center}
  \caption{The renewal effect}
  \label{RZipf}
\end{figure}

In this paper we assume that the document rate of change $\mu(i)$
depends on its popularity $i$. The mathematical equivalent of this
assertion is that the steady-state process is again described by
Zipf-like distribution with a small $\alpha_R$ as it is shown of
the Fig.~\ref{RZipf}.

The difference between an ideal performance of a cache system and
its real value has been conditioned by the renewal of documents in
the global network
\begin{equation}\label{deltaH}
  \triangle H=\left(
  \sum\limits_{i=1}^M \vartheta_i-HK\right)/K
\end{equation}
and can be investigated in an experimental way. Here
\begin{equation}\label{dif-k}
\triangle k=\sum\limits_{i=1}^M\vartheta_i-HK=k-HK+M-p,
\end{equation}
$\triangle k$ is the number of updating requests conditioned by
renewal of Web documents and
\begin{equation}
 k_R =
HK+p-M \label{k-R}
\end{equation}

Therefore we have to explore the {\em log\/} file collected for
the time $T_{st}\lesssim T_{eff}-\triangle T$ using the data from
the Tab.~\ref{SR}. Then the condition $M(T_{st})\lesssim S_k$ has been
fulfilled and all repeated requests are explained by the renewal
effect. The ideal Zipf-like distribution corresponds to the upper
line on the Fig.~\ref{RZipf} with exponent $\alpha$ from
Eq.~(\ref{alpha-p}), but the real hit ratio $H$ determines
$\alpha_R$ as
\begin{equation}\label{alR}
  \alpha_R=1-2M/HK
\end{equation}
Now it is easy to find $\mu_i$
\begin{equation}\label{mui}
  \mu(i)=\frac{\vartheta^{id}_i-\vartheta^R_i}{T_{st}}=
  \frac{(p/i)^\alpha-(p/i)^{\alpha_R}}{T_{st}},
\end{equation}
or
\begin{equation}\label{rasp-f}
  \mu(i)=\frac{1}{T_{st}}
  \frac{1-(i/p)^{\triangle\alpha}}{(i/p)^{\alpha}}.
\end{equation}
We can consider $\mu_u=\mu(p/4)$ and $\mu_p =\mu(p/100)$, then the
results can be summarized in the Table~\ref{al-R-e}.

\begin{table*}
\caption{The renewal parameters}\label{al-R-e}{\scriptsize
\begin{tabular}{|c|c|c|c|c|c|c|}
\hline $\alpha$ & $\alpha_R$ & $\triangle H $ & $H$& $\mu_p$ &
$\mu_u$ & $T_{st}$ \\ \hline 0.72 & 0.7 & 2.3\% & 32.04\% & 1/6.2
days & 1/202 days & 15 days
\\ \hline
\end{tabular}}
\end{table*}

The key difference between our analytical model and presented
by Wolman at al. model are:
\begin{itemize}
\item
The document rate of change $\mu(i)$ is described by continuous
variables that depend on popularity $i$. Wolman et al. assume that
the document rate of change $\mu$ takes two different values, one
for popular documents $\mu_p$ and another for unpopular documents
$\mu_u$.
\item
The renewal of Web documents is incorporated in a steady-state
process as a Zipf-like distribution with a small $\alpha_R$.
\item
We assume also that size of cache system is restricted.
\item
A considerable part of cacheable documents $p-M$ are requested
from the global network only one time accordingly to Zipf-like
distribution. The more accurate expression for an ideal hit ratio
$H_i$ has been shown in Ref.~\cite{dols1}
\begin{equation}\label{H-i-a}
  H_i\leq 2^{(\alpha-1)/\alpha},
\end{equation}
It has to be modified to taking into account the renewal effect:
\begin{equation}\label{H_ren}
  H_i\leq 2^{(\alpha-1)/\alpha}(1-\alpha)/(1-\alpha_R),
\end{equation}
\end{itemize}

Two types of correlations pretend to a role of fundamental laws
that describe cache systems~\cite{dols}:
\begin{itemize}
  \item A Zipf-like distribution, see Eq.~(\ref{zipf-gen})
  \item Normalizing conditions or a sum of the
  probability to request the universe of $1 \leq n \leq k$ objects.
\end{itemize}

The above mentioned laws could be applied to the special points of
Zipf distribution and two of them, $M$ and $p$, are used for the
construction of the theory. Then the Zipf-like distribution leads
to
\begin{eqnarray}
\label{A_M} \frac{Ak_R}{M^{\alpha_R}} &= & 2,\\ \label{A_p}
\frac{Ak_R}{p^{\alpha_R}} & =& 1.
\end{eqnarray}
Normalizing conditions for the first $M$ and $p$ documents from
cache give
\begin{eqnarray}
\label{Hi-n} \int\limits_1^M \frac{A}{x^{\alpha_R}} dx &=& H_i\\
\label{1-n} \int\limits^p_1\frac{A}{x^{\alpha_R}} dx &=& 1
\end{eqnarray}
Here $H_i$ is an ideal (steady-state) performance for cacheable
documents. For a real system the Eq.~(\ref{Hi-n}) has been
transformed to
\begin{equation}\label{H_rn}
 H=p_c\int\limits_1^{S_k}\frac{A}{x^{\alpha_R}}dx,
\end{equation}
where $S_k$ is the number of cache objects in the kernel.

It is important that Zipf exponent $\alpha$ grows as the
experiment time $T_{st}$ increases.
\begin{equation}\label{inc-al}
  \triangle
  \alpha=\alpha(T^2_{st})-\alpha(T^1_{st})=C\ln(T^2_{st}/T^1_{st})
\end{equation}

\section{Summary and future work}
\label{work}

The aim of this paper is a study of a Web cache system in order to
optimize proxy cache systems and to modernize construction
principles. Our investigations lead to the criteria for the
optimal usage of storage capacity and to allow to description of
the basic effects as the ratio between the construction parts,
steady-state performance, optimal size, etc. We want to outline
that the results received and the plan of the experiment follow
from the theoretical model.

Special consideration is given to the modification of the key
formulas supposed by Wolman at al.~\cite{wolm}. The document rate
of change $\mu(i)$ is supposed to depend on popularity index $i$
so the Zipf-like distribution with the new exponent $\alpha_R$
describes the effects of the renewal of Web documents.

The main result of any research of cache system is finding a way
for increasing hit ratio. We can conclude that with growth of
$S_k/S_u$ the hit rate $H$ increases. A general feature of the
current algorithm is the only documents requested two times and
more during the time $t_u$ are included in the kernel $S_k$. It is
a fact that $M$ less then the cache size $S_{eff}$, i.e. all
necessary items could be stored in a cache system at the second,
third and forth experimental points on the Tab.~\ref{SR}.
Therefore we can make the conclusion that a caching algorithm
based on a rigid tie of the mean parameters to the time $t_u$ is
ineffective.

One method of resolving the current situation is reconstructing of
cache systems. Such a construction suggested in the
Sec.~\ref{correl} must be implemented to provide a rigid ratio
between its elements. The requests' statistics must be kept even
for those cacheable documents that have been deleted from the
cache. This statistics should exist for long time. This is one of
the key differences between our construction and the existing ones
that usually operate only with the current set of documents in
cache. An inseparable part of the new construction is the
replacement algorithm based on Zipf law. Now we have prepared the
corresponding application for an international patent.

An additional question, which will demanded a new experimental
search, is the dependence that describes the renewal of Web
documents. Probably, the ratio $\triangle\alpha/\alpha$ could be
considered as a universal constant. The next direction for our
development plans is an investigation of a model of cache
interaction in hierarchical caching system.

\end{document}